\documentclass{article}
\usepackage{nips10submit_e,times}

\usepackage[dvips]{graphicx}
\usepackage{subfig}

\usepackage[numbers]{natbib}

\newcommand{\be}{\begin{equation}}
\newcommand{\en}{\end{equation}}
\newcommand{\naive}{na\"{\i}ve }

\title{Rescaling, thinning or complementing? On goodness-of-fit procedures for point process models and Generalized Linear Models}

\author{
Felipe Gerhard \\
Brain Mind Institute\\
Ecole Polytechnique F\'{e}d\'{e}rale de Lausanne\\
1015 Lausanne EPFL, Switzerland\\
\texttt{felipe.gerhard@epfl.ch} \\
\And
Wulfram Gerstner \\
Brain Mind Institute\\
Ecole Polytechnique F\'{e}d\'{e}rale de Lausanne\\
1015 Lausanne EPFL, Switzerland\\
\texttt{wulfram.gerstner@epfl.ch} \\
}

\nipsfinalcopy 

\begin{document}

\maketitle

\begin{abstract}
Generalized Linear Models (GLMs) are an increasingly popular framework for modeling neural spike trains. They have been linked to the theory of stochastic point processes and researchers have used this relation to assess goodness-of-fit  using methods from point-process theory, e.g. the time-rescaling theorem. However, high neural firing rates or coarse discretization lead to a breakdown of the assumptions necessary for this connection. Here, we show how goodness-of-fit tests from point-process theory can still be applied to GLMs by constructing equivalent surrogate point processes out of time-series observations. Furthermore, two additional tests based on thinning and complementing point processes are introduced. They augment the instruments available for checking model adequacy of point processes as well as discretized models.
\end{abstract}

\section{Introduction}

Action potentials are stereotyped all-or-nothing events, meaning that their amplitude is not considered to transmit any information and only the exact time of occurrence matters. This view suggests to model neurons' responses in the mathematical framework of point processes. An observation is a sequence of spike times and their stochastic properties are captured by a single function, the conditional intensity \cite{Daley2003}. For point processes on the time line, several approaches for evaluating goodness-of-fit have been proposed \cite{Ogata1981}. The most popular in the neuroscientific community has been a test based on the time-rescaling theorem \cite{Brown2002}.

In practice, neural data is binned such that a spike train is represented as a sequence of spike counts per time bin. Specifically, Generalized Linear Models (GLMs) are built on this representation. Such discretized models of time series have mostly been seen as an approximation to continuous point processes and hence, the time-rescaling theorem was also applied to such models \cite{Barbieri2001,Koyama2008,Rigat2006,Shimokawa2009,Wojcik2009}.

Here we ask the question whether the time-rescaling theorem can be translated to discrete time. We review the approximations necessary for the transition to discrete time and point out a procedure to create surrogate point processes even when these approximations do not hold (section~\ref{sec:methods}). Two novel tests based on two different operations on point processes are introduced: random thinning and random complementing. These ideas are applied to a series of examples (section~\ref{sec:results}), followed by a discussion (section~\ref{sec:discussion}).

\section{Methods}
\label{sec:methods}

\subsection{Representations of neural activity}
\label{sec:representationofneuralactivity}

\begin{figure}
\begin{center}
 \includegraphics[width=0.99\linewidth]{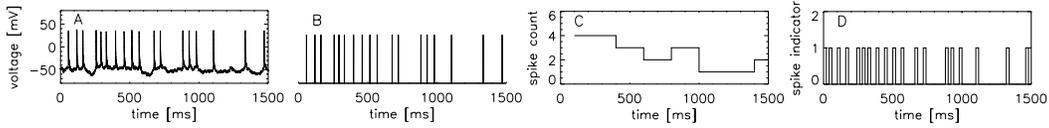}
\end{center}
 \caption{Spike train representations. (A) A trace of the membrane potential of a spiking neuron. (B) Information is conveyed in the timings and number of action potentials. This supports the representation of neural activity as a point process in which each spike is assumed to be a singular event in time. (C) When time is divided into large bins, the spike train is represented as a time series of discrete counts. (D) If the bin width is chosen small enough, the spike train corresponds to a binary time series, indicating the presence of a single spike inside a given time bin.}
\label{fig:spiketrainrepresentations}
\end{figure}

We characterize a neuron by its response in terms of trains of action potentials using the theory of \emph{point processes} (Figures~\ref{fig:spiketrainrepresentations}A and \ref{fig:spiketrainrepresentations}B). An observation consists of a list of times, each denoting the time point of one action potential. Following a common notation \cite{Brown2002,Truccolo2005}, let $(0,T]$ be the time interval of the measurement and $\{u_i\}$ be the set of $n$ event times. The stochastic properties of a point process are characterized by its conditional intensity function $\lambda(t|H(t))$, defined as \cite{Daley2003}:

\begin{equation}
\label{eq:conditionalintensity}
    \lambda(t|H_t) = \lim_{\Delta \rightarrow 0} \frac{P[\mathrm{spike\ in\ }(t,t+\Delta)|H_t]}{\Delta} ,
\end{equation}
where $H_t$ is the history of the stochastic process up to time $t$ and possibly includes other covariates of interest. For fitting and evaluating different parameter sets of the conditional intensity function, a maximum-likelihood approach is followed \cite{Pawitan2001,DoyaBook2007}. The log-likelihood of a point process model is given by \cite{Daley2003}:

\begin{equation}
\label{eq:likelihoodforcond}
    \log L(\mathrm{point\ process}) = \sum_{i=1}^n \log \lambda(u_i|H_{u_i}) - \int_0^T \lambda(t|H_t)dt .
\end{equation}

One possibility are binning-free models (like renewal processes or other parametric models). Alternatively, $\lambda(t|H_t)$ can be modeled as a piece-wise constant function with each piece having length $\Delta$. In this case, the history term $H_t$ covers the history up to the time of the left edge of the current bin. Inside the bin, the process locally behaves like a Poisson process with constant rate $\lambda_k = \lambda(t_k|H_k)$ with $t_k = \Delta k$ and $H_k = H_{t_k}$. Using the number of spikes $c_k$ per bin as a representation of the observation, the discretized version of Equation~\ref{eq:likelihoodforcond} is equivalent to the log-likelihood of a series of Poisson samples (apart from terms that are not dependent on $\lambda(t|H_t)$). Hence, for finding the maximum-likelihood solution for the point process, it is equivalently sufficient to maximize the likelihood of such a Poisson regression model. The result of fitting will be a sequence of $\mu_i$ for each bin, where $\mu_i$ is the expected number of counts. Since a local Poisson process is assumed within the bins, $\mu_i$ is related to $\lambda_i$ via: $\lambda_i = \mu_i/\Delta$.

A complementary approach to the point process framework is to see spike trains as \textit{time series}, e.~g.~as a sequence of counts $\{c_i\}$ or binary events $\{b_i\}$ (Figures~\ref{fig:spiketrainrepresentations}C and \ref{fig:spiketrainrepresentations}D). For Poisson-GLMs, a sequence of Poisson-distributed count variables $c_i$ is modeled and the linear sum of covariates is linked to the expected mean of the Poisson distribution $\mu_i$.
Binary time series can be modeled as a sequence of conditionally independent Bernoulli trials with outcomes 0 and 1 and success probabilities $\{p_i\}$. For Bernoulli-GLMs, the $p_i$s are linked via a non-linear transfer function to a linear sum of covariates. Defined this way, the likelihood for an observed sequence $b_i$ given a particular model of $p_i$ is given by $\log L(\mathrm{Bernoulli}) = \sum_k b_k \log \frac{p_k}{1-p_k} + \sum_k \log(1-p_k)$. In the approximation of $\mu_i \ll 1$, $\mu_i$ becomes approximately $p_i$ and the likelihoods of the Bernoulli and Poisson series become equivalent. Moreover, using the same approximation, it is possible to link the Bernoulli series to the conditional intensity function $\lambda(t|H_t)$ via $\lambda_i \approx p_i/\Delta$ . Traditionally, this path was chosen to relate the time series to the theory of point processes and to be able to use goodness-of-fit analyses available for such point processes \cite{Truccolo2005}.

\subsection{Goodness-of-fit tests for point processes}

Statistical tests are usually evaluated using two measures: The \emph{specificity} (fraction of correct models that pass the test) and the \emph{sensitivity} or \emph{test power} (fraction of wrong models that are properly rejected by the test). The specificity is set by the significance level: With significance level $\alpha$, the specificity is $1-\alpha$. The sensitivity of a given test depends on the strength of the departure from the modeled intensity function to the true intensity.

\subsubsection{The time-rescaling theorem}

A popular way for verifying point-process-based models has been the time-rescaling theorem \cite{Brown2002,Pillow2009}. It states that if $\{u_i\}$ is a realization of events from a point process with conditional intensity $\lambda(t|H_t)$, then rescaling via the transformation $ u_i' = \int_0^{u_i} \lambda(t|H_t) dt $ will yield a unit-rate Poisson process.

We call the following transformation the \emph{na\"{\i}ve time-rescaling} when it is applied to binary sequences. The spike time $u_i$ falling into bin $j$, is transformed into: $u_i' = \sum_{k=1}^{j} p_k$.

\subsubsection{Thinning point processes}
\label{sec:gof_thinning}

\begin{figure}
\begin{center}
 \includegraphics[width=0.90\linewidth]{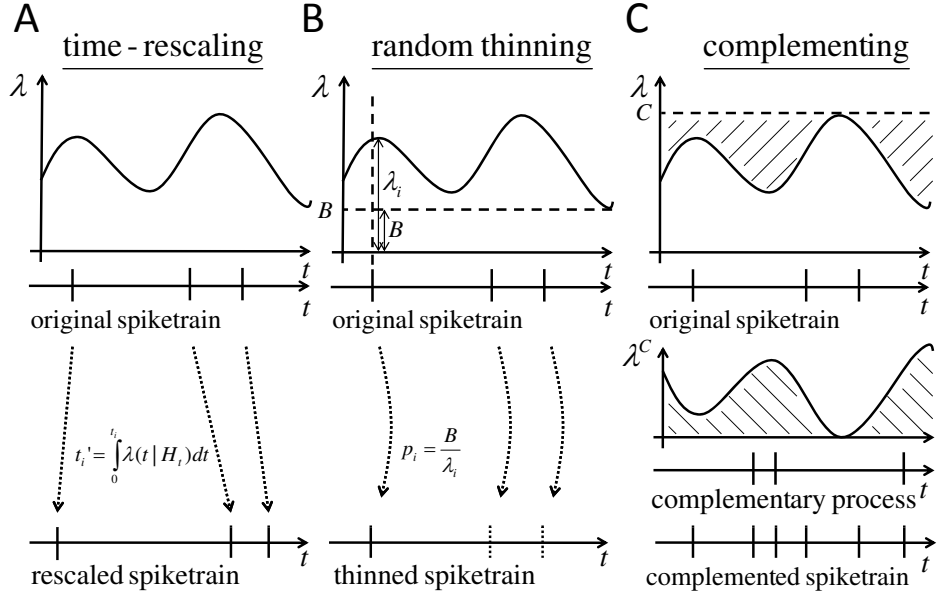}
\end{center}
 \caption{Overview of goodness-of-fit tests for point-process models. (A) Using the time-rescaling theorem, the time of each spike is rescaled according to the integral of the conditional intensity function. (B) Assuming that the conditional intensity function has a lower limit $B$, spikes of the original spike train are thinned by keeping a spike only with probability $B \lambda_i^{-1}$. (C) Assuming that the conditional intensity function has an upper limit $C$, a complementary process $\lambda^C = C-\lambda$ can be constructed. Adding samples from this inhomogeneous Poisson process to the observed spikes results in a homogeneous Poisson process with rate $C$.}
\label{fig:scheme}
\end{figure}

It is well known that an inhomogeneous point process can be simulated by generating a homogeneous Poisson process with constant intensity $C$ with $C \geq \max \lambda(t)$ (the so-called dominant process) and keeping every spike at time $t_i$ with probability $p = \frac{\lambda(t_i)}{C}$ \cite{Lewis1979,Ogata1981}. In reverse, this can be used to do model-checking \cite{Schoenberg2003}: Let $B$ be a lower bound of the fitted conditional intensity $\lambda(t|H(t))$. Now take $\lambda(t|H(t))$ as the dominant process with samples $u_i$. Thin the process by keeping a spike with probability $\frac{B}{\lambda(t_i|H_t)}$. For a correctly specified model $\lambda(t|H_t)$, the thinned process will be a homogeneous Poisson process with rate $B$ (Figure~\ref{fig:scheme}B).

Typically, $B = \min \lambda(t) \ll \bar{\lambda(t)}$ (due to absolute refractoriness in most renewal process models and GLMs), such that the thinned process will have a prohibitively low rate and only very few spikes will be selected. Testing the Poisson hypothesis on a handful of spikes will result in a vanishingly low power.

To circumvent this problem, we propose the following remedy: Let $B^*$ be a threshold which may be higher than the lower bound $B$. Then consider only the intervals of $\lambda$ for which $\lambda > B^*$ and concatenate those into a new point process. After applying the thinning procedure on all spikes of the stitched process, the thinned process should be a Poisson process with rate $B^*$. This procedure can be repeated $K$ times for a range of uniformly spaced $B^*$s ranging from $B$ to $C$ (upper bound). Stretching each thinned process by a factor of $B^*$ creates a set of $K$ unit-rate processes. Each of them is tested for the Poisson hypothesis by a Kolmogorov-Smirnov test on the inter-spike intervals. The model is rejected when there is at least one significant rejected null hypothesis. To correct for the multiple tests, we employ Simes' procedure. It tests the global null hypothesis that all tested sub-hypotheses are true against the alternative hypothesis that at least one hypothesis is false. To this end, it transforms the ordered list of p-values $p^{(1)}, ..., p^{(K)}$ into $\frac{K p^{(1)}}{1}, \frac{K p^{(2)}}{2}, ..., \frac{K p^{(K)}}{K}$. If any of the transformed p-values is less than the significance level $\alpha = .05$, the model is rejected \cite{Simes1986}\footnote{The $K$ tests contain overlapping regions of the same spike train, hence, we expect the statistical tests to be correlated. In these cases, a simple Bonferroni-correction would be too conservative \cite{Rodland2006}.}.

\subsubsection{Complementing point processes}
\label{sec:gof_complementing}

The idea of thinning might also be used the other way round. Assume the observations $u_i$ have been generated by thinning a homogeneous Poisson process with rate $C$ using the modeled conditional intensity $\lambda(t|H_t)$ as the lower bound. Then we can define a complementary process $\lambda^{c}(t) = C-\lambda(t|H_t)$ such that adding spikes from the complementary point process to the observed spikes, the resulting process will be a homogeneous Poisson process with rate C. This algorithm is a straight-forward inversion of the thinning algorithms discussed in \cite{Ogata1981,Daley2003}.

It might happen that the upper bound $C$ of the modeled intensity is much larger than the average $\lambda(t)$. In that case, the observed spike pattern would be distorted with high number of Poisson spikes from the complementary process and the test power would be low. To avoid this, a similar technique as for the thinning procedure can be employed. Define a threshold $C^* \leq C$ and consider only the region of the spike train for which $\lambda(t|H(t)) < C^{*}$. Apply the complementing procedure on these parts of the spike train to obtain a point process with rate $C^{*}$ when concatenating the intervals. This process can be repeated $K$ times with values $C^{*}$ ranging from $B$ to $C$. A multiple-test correction has to be used, again we propose Simes' method (see previous section).

\subsection{Creating surrogate point processes from time series}
\label{sec:surrogate}

\newcommand{\exdiscrate}{50 }

Since the time-rescaling theorem can only be used when $\lambda(t|H_t)$ the exact spike times $\{u_i\}$ are known, it is not a priori clear how it applies to discretized time-series models. For such cases, we propose to generate surrogate point process samples that are equivalent to the observed time series. To apply the time-rescaling theorem on discretized models such as GLMs, the integral of the time transformation is replaced by a discrete sum over bins (the \emph{\naive time-rescaling}). Taking the simplest example of a homogeneous Poisson process, it is evident that the possible values for the rescaled intervals form a finite set. This contradicts the time-rescaling theorem that states that the intervals are (continuously) exponentially distributed. Hence, using the time-rescaling theorem on discretized data produces a bias \cite{Haslinger2010DTR}.

\begin{figure}
\begin{center}
 \includegraphics[width=1\linewidth]{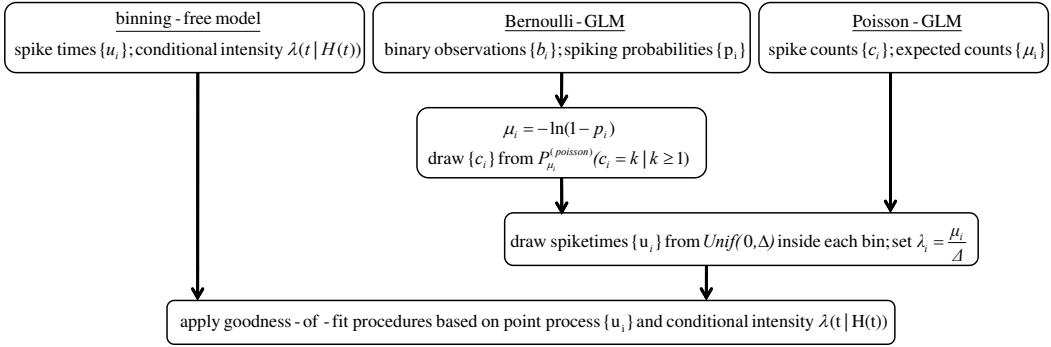}
\end{center}
 \caption{Creating surrogate point processes from time series. For bin-free point process models for which the spike times and a conditional intensity $\lambda(t|H(t))$ is available, goodness-of-fit tests for point processes can be readily applied. For Poisson-GLMs, exact spike times are drawn inside each bin for the specified number of spikes that were observed. The piece-wise constant conditional intensity function is linked to the modeled number of counts per bin via $\lambda_i = \Delta^{-1} \mu_i$. For Bernoulli-GLMs, the probability of obtaining at least one spike per bin $p_i$ is modeled. For each bin with spikes ($b_i = 1$) -- assuming a local Poisson process -- a sample $c_i$ from a biased Poisson distribution with mean $\mu_i = -\ln(1-p_i)$ is drawn together with corresponding spike times. Finally, point-process based goodness-of-fit tests may be applied to this surrogate spike train.}
\label{fig:flowchart}
\end{figure}

While Haslinger et al.~considered a modification of the time-rescaling theorem to explicitly account for the discrete nature of the model \cite{Haslinger2010DTR}, we propose a general, simple scheme how to form surrogate point processes from Poisson- and Bernoulli-GLMs that can be used for the continuous time-rescaling theorem as well as for any other goodness-of-fit test designed for point-process data (Figure~\ref{fig:flowchart}). 


\textbf{Poisson-GLMs}: The observation consists of a sequence of count variables $c_i$ that is modeled as a sample from Poisson distributions with mean $\mu_i$. Hence, the modeled process can be regarded as a piecewise-constant intensity function. The expected number of spikes of a Poisson process is related to its intensity via $\mu_i = \lambda_i \Delta$ such that we can construct the conditional intensity function as piece-wise constant with values $\lambda_i = \Delta^{-1}{\mu_i}$. Conditioned on the number of spikes that occurred in a homogeneous Poisson process of rate $\lambda_i$, the exact spike times are uniformly distributed inside bin $i$. A surrogate point process can be constructed from a Poisson-GLM by generating random spike times $(i - 1 + Unif(0,1)) \Delta$ for each spike within bin i ($1 \le i \le N$) for all bins with $c_i > 0$. One can then proceed to the point-process-based goodness-of-fit tools using the surrogate spike train and its conditional intensity $\lambda_i$.


\textbf{Bernoulli-GLMs}: Based on the observed binary spike train $\{b_i\}$, the sequence of probabilities $p_i$ of spiking within bin $i$ is modeled. We can relate this to the point process framework using the following observations: Assume that $p_i$ denotes the probability of finding at least one spike within each bin\footnote{Such clipping is implicitly performed in many studies, e.~g.~in \cite{Schneidman2006,Pillow2008b,Tang2008}.} and that locally, the process behaves like a Poisson process. Then, $p_i = P_{\mu_i}^{\rm (poisson)}(X \ge 1) = 1 - P_{\mu_i}^{\rm (poisson)}(X = 0) = 1-\exp(-\mu_i)$. The conditional intensity is given by $\lambda_i = \Delta^{-1}{\mu_i} = - \Delta^{-1} \ln(1-p_i)$. In practice, for each bin with $b_i = 1$, we draw the amount of spikes within the bin by first sampling from the distribution $P_{\mu_i}^{\rm (poisson)}(X=k|k \ge 1)$ and sample exact spike times uniformly as in the case of the Poisson-GLMs.

\section{Results}
\label{sec:results}

Here, we compare the performance of the three different approaches in detecting wrongly specified models, using examples of models that are commonly applied in neural data analysis. For the thinning and complementing procedure, $K = 10$ partitions were chosen (see section~\ref{sec:gof_thinning}). Unless otherwise noted, we report the test power at a specificity of $1 - \alpha = .95$. The Poisson hypothesis in the proposed procedures is tested by a Kolmogorov-Smirnov test on the inter-spike intervals of the transformed process.

\subsection{Example: Inhomogeneous Poisson process}

\newcommand{\exinhomoncoeff}{40 }
\newcommand{\exinhomobandlimit}{1 } 
\newcommand{\exinhomoT}{20 } 
\newcommand{\exinhomoDelta}{1 } 
\newcommand{\exinhomoNrep}{1000 }
\newcommand{\exinhomomedjitter}{12 }
\newcommand{\exinhomolargejitter}{30 }

Consider an inhomogeneous Poisson process with band-limited intensity: $ \lambda(t|H_t) = \lambda(t) = 20~\mathrm{Hz} + \sum_{j=1}^{J = \exinhomoncoeff} u_j \frac{sin(2 \pi f (t-\frac{j}{J}T))}{\pi (t-\frac{j}{J}T)} $ with $f = \exinhomobandlimit$~Hz and $J = \exinhomoncoeff$ coefficients that were randomly drawn from a uniform distribution on the interval $[0,20]$. The process was simulated over a length of $T = \exinhomoT$~s and the intensity was discretized with $\Delta = \exinhomoDelta$~ms. Negative intensities were clipped to zero. A binary spike train was generated by calculating the probability of at least one spike in each time bin as $p_i = 1-\exp(-\lambda(t_i) \Delta)$ and drawing samples from a Bernoulli distribution with specified probabilities $p_i$.

\begin{figure}
\centering
 \subfloat[intensity function]{\includegraphics[width=0.33\linewidth]{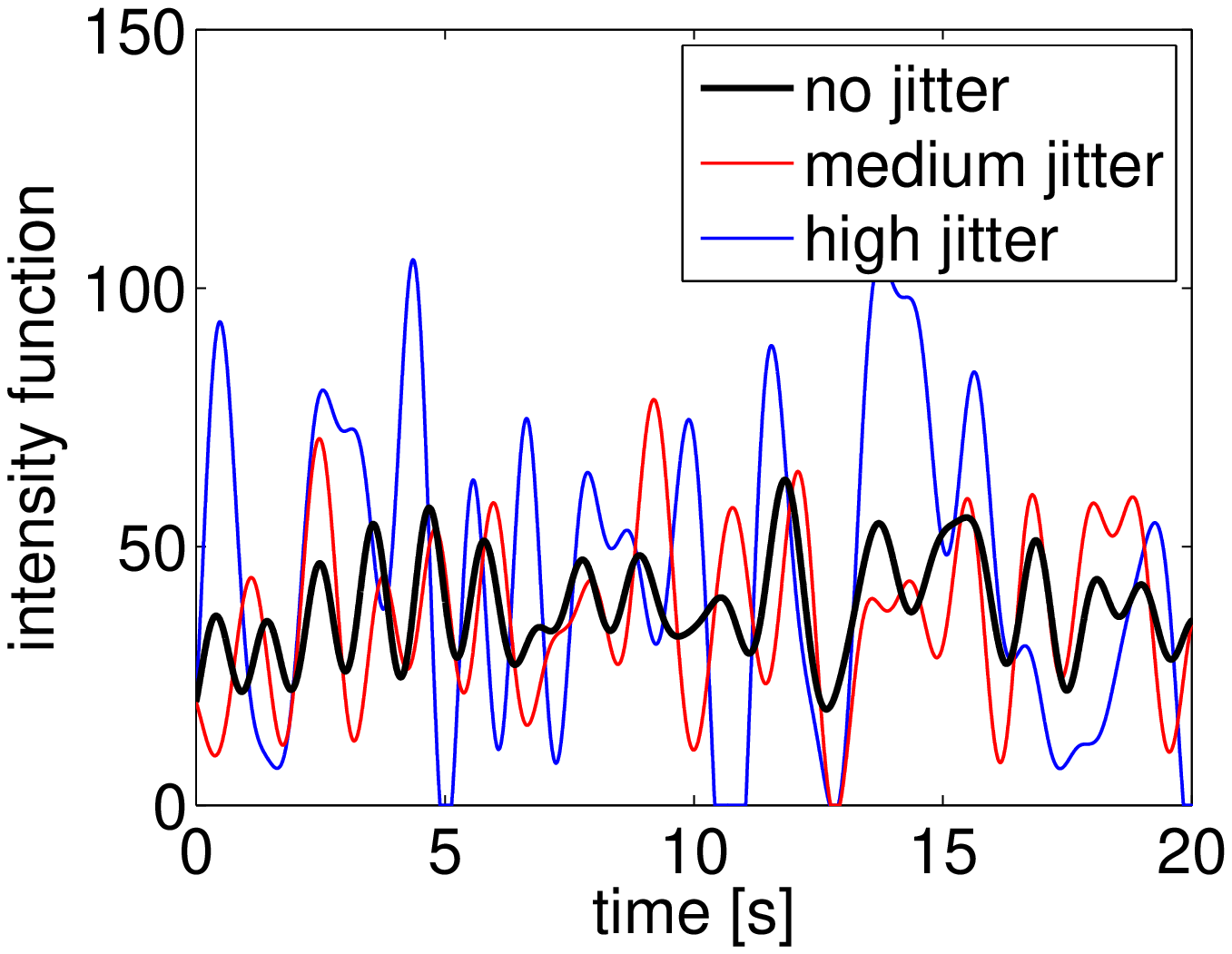}}
 \subfloat[test power]{\includegraphics[width=0.33\linewidth]{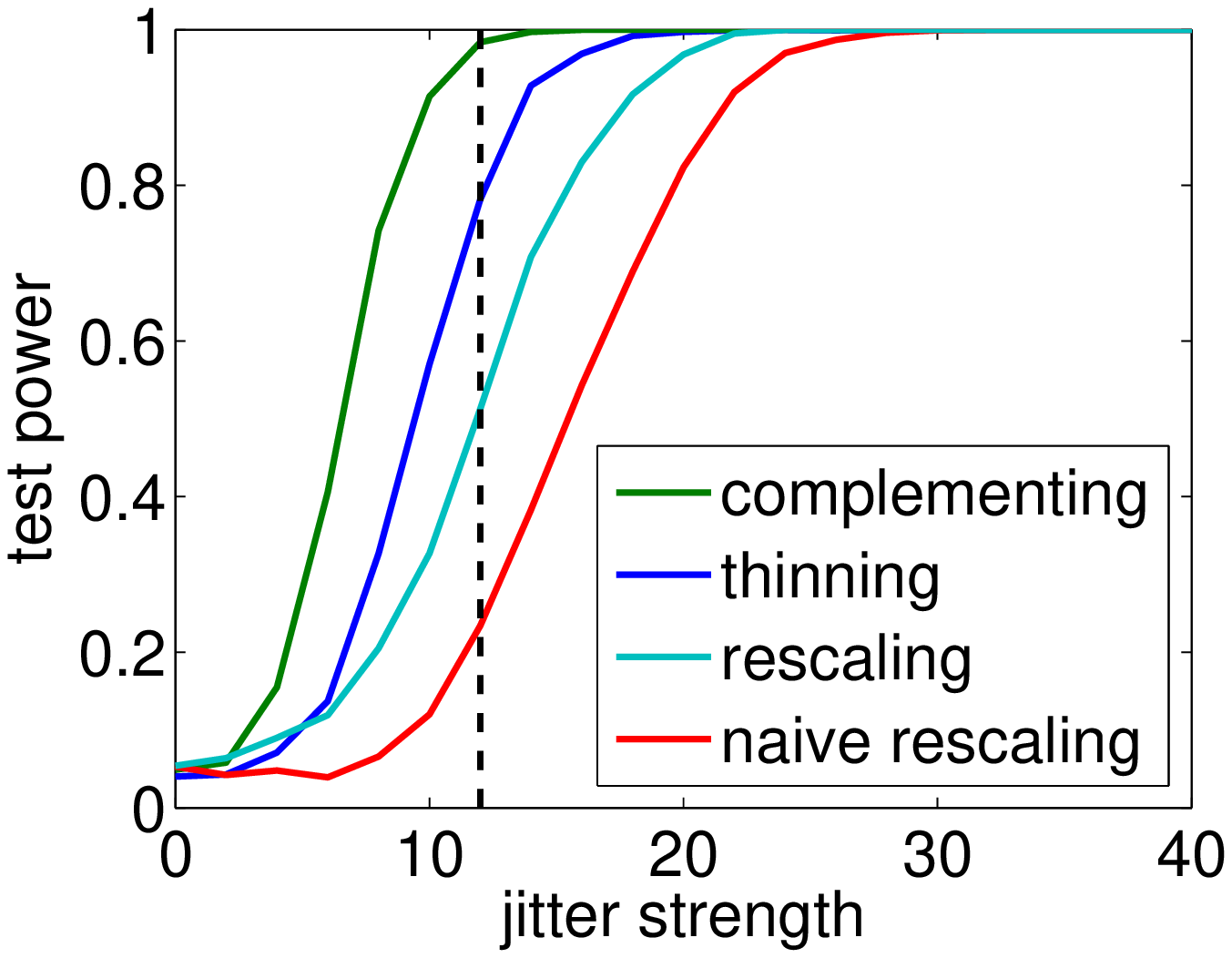}}
 \subfloat[ROC curve]{\includegraphics[width=0.33\linewidth]{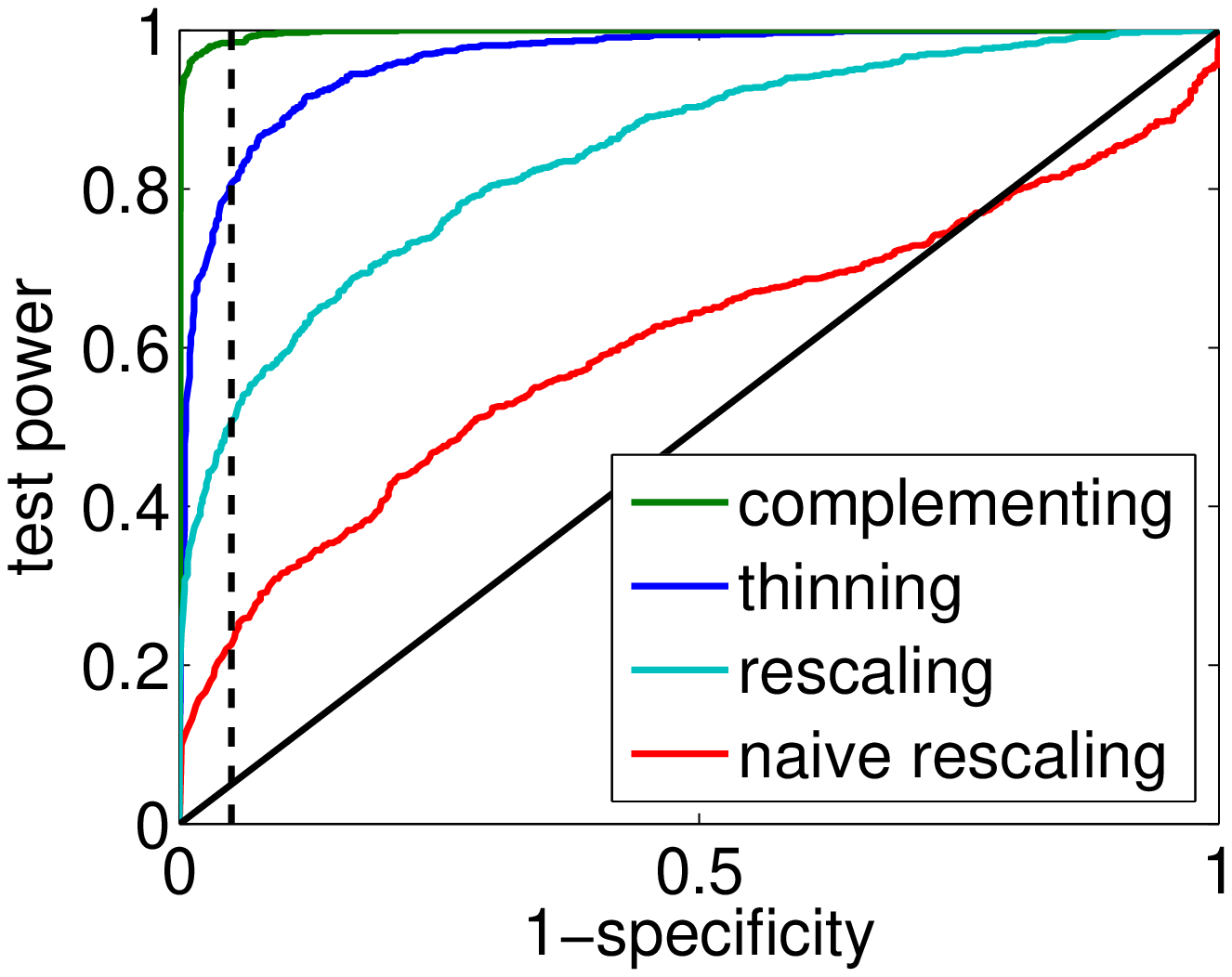}}
 \caption{Inhomogeneous Poisson process. (A) Sample intensity functions for an undistorted intensity (black line) and two models with jitters in the coefficients ($\beta = \exinhomomedjitter$, medium jitter and $\beta = \exinhomolargejitter$, large jitter). (B) The test power of each test as a function of the jitter strength. The dashed line indicates the level of the medium jitter strength (red line in figure A). (C) ROC curve analysis for an intermediate jitter strength of $\beta = \exinhomomedjitter$. The intersection of the curves with the dashed line corresponds to the test power at a significance level of $\alpha = .05$.}
\label{fig:exinhomo}
\end{figure}

For evaluating the different algorithms, wrong models for the intensity were created with jittered coefficients $u'_k = u_k + \beta \mathrm{Unif}(-1,1)$ where $\beta$ indicates the strength of the deviation from the true model. For each jitter strength, $N = \exinhomoNrep$ spike trains were generated from the true model and $\lambda(t|H_t)$ was constructed using the wrong model (Figure~\ref{fig:exinhomo}A). For any $\beta > 0$, the fraction of rejected models defines the sensitivity or test power. For $\beta = 0$, the fraction of accepted models defines the specificity which was controlled to be at $1-\alpha = .95$ for each test.

All three methods (rescaling, thinning, complementing) show a specified type-I error of approximately 5\% ($\beta = 0$) and progressively detect the wrong models. Notably, the complementing and thinning procedures detect a departure from the correct model earlier than the classical rescaling (Figure~\ref{fig:exinhomo}B). For comparison, also the \naive implementation of the rescaling transformation is shown. The significance level for the KS test used for the \naive time-rescaling was adjusted to $\alpha = .015$ to achieve a 95\% specificity. The adjustment was necessary due to the discretization bias (see section~\ref{sec:surrogate}).

For models with an intermediate jitter strength ($\beta = \exinhomomedjitter$), ROC curves were constructed. Here, for a given significance level $\alpha$, a pair of true and false positive rates can be calculated and plotted for each test (taking $N = \exinhomoNrep$ repetitions using the true model and the model with jittered coefficients). It can be seen that especially for intermediate jitter strengths, complementing and thinning outperform time-rescaling (Figure~\ref{fig:exinhomo}C), independent of the chosen significance level.

\subsection{Example: Renewal process}

\newcommand{\exgammaT }{20 } 
\newcommand{\exgammaDelta}{1 } 
\newcommand{\exgammaNrep}{1000 }
\newcommand{\exgammatruescale}{0.032 }
\newcommand{\exgammatrueshape}{6.25 }
\newcommand{\exgammamedjitter}{0.5 }
\newcommand{\exgammahighjitter}{1.0 }

In a second example, we consider renewal processes, i.~e.~inter-spike intervals are an i.~i.~d. sample from a specific probability distribution $p(\Delta t)$. In this case, the conditional intensity is given by $\lambda(t|H_t) = \frac{p(t-t^{*})}{1-\int_0^{t-t^{*}} p(u) du}$ where $t^{*}$ denotes the time of the last spike prior to time t. For this example, we chose the Gamma distribution as it is commonly used to model real spike trains \cite{Barbieri2001,Brown2002,Shimokawa2009}.

\begin{figure}
\centering
 \subfloat[intensity function]{\includegraphics[width=0.33\linewidth]{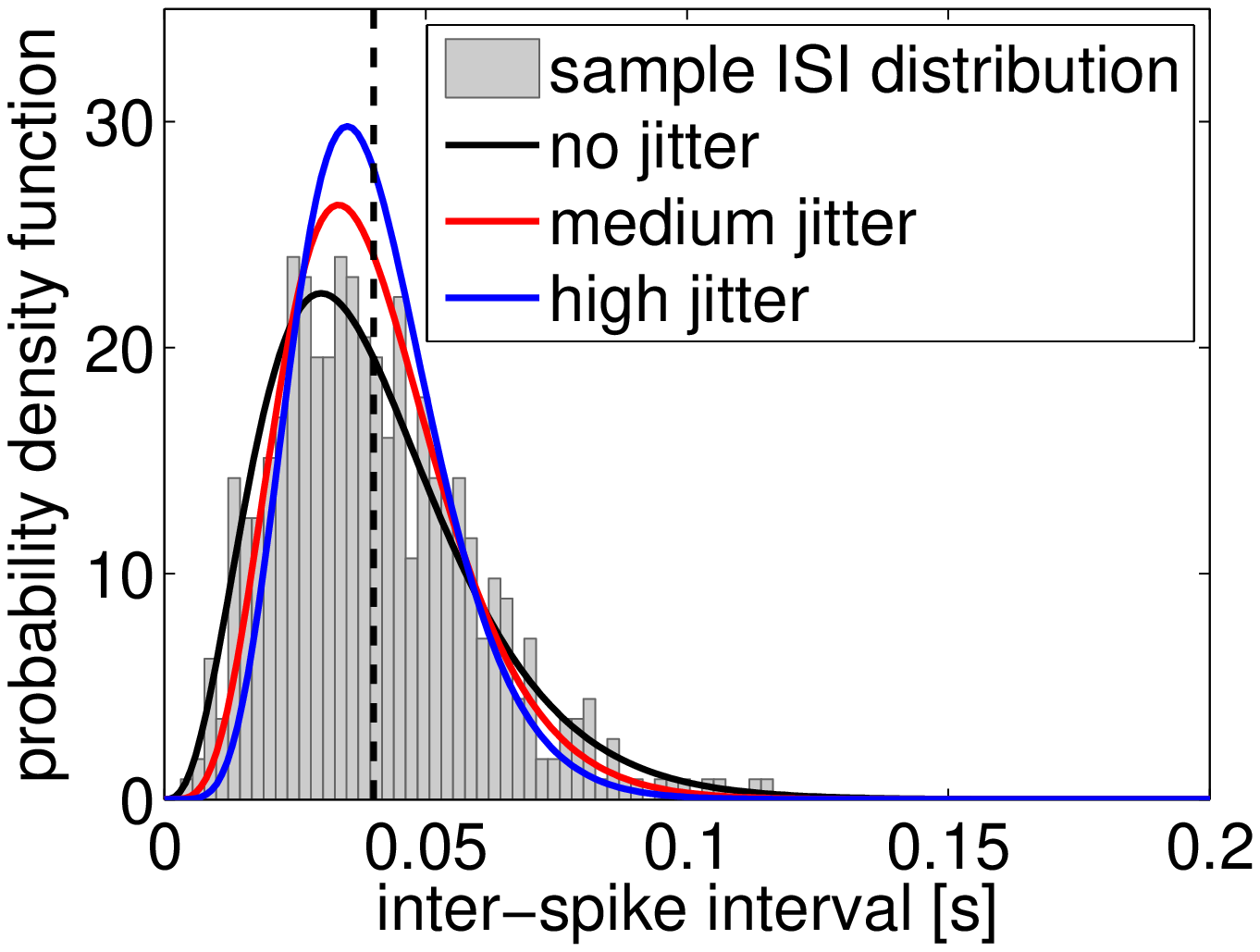}}
 \subfloat[test power]{\includegraphics[width=0.33\linewidth]{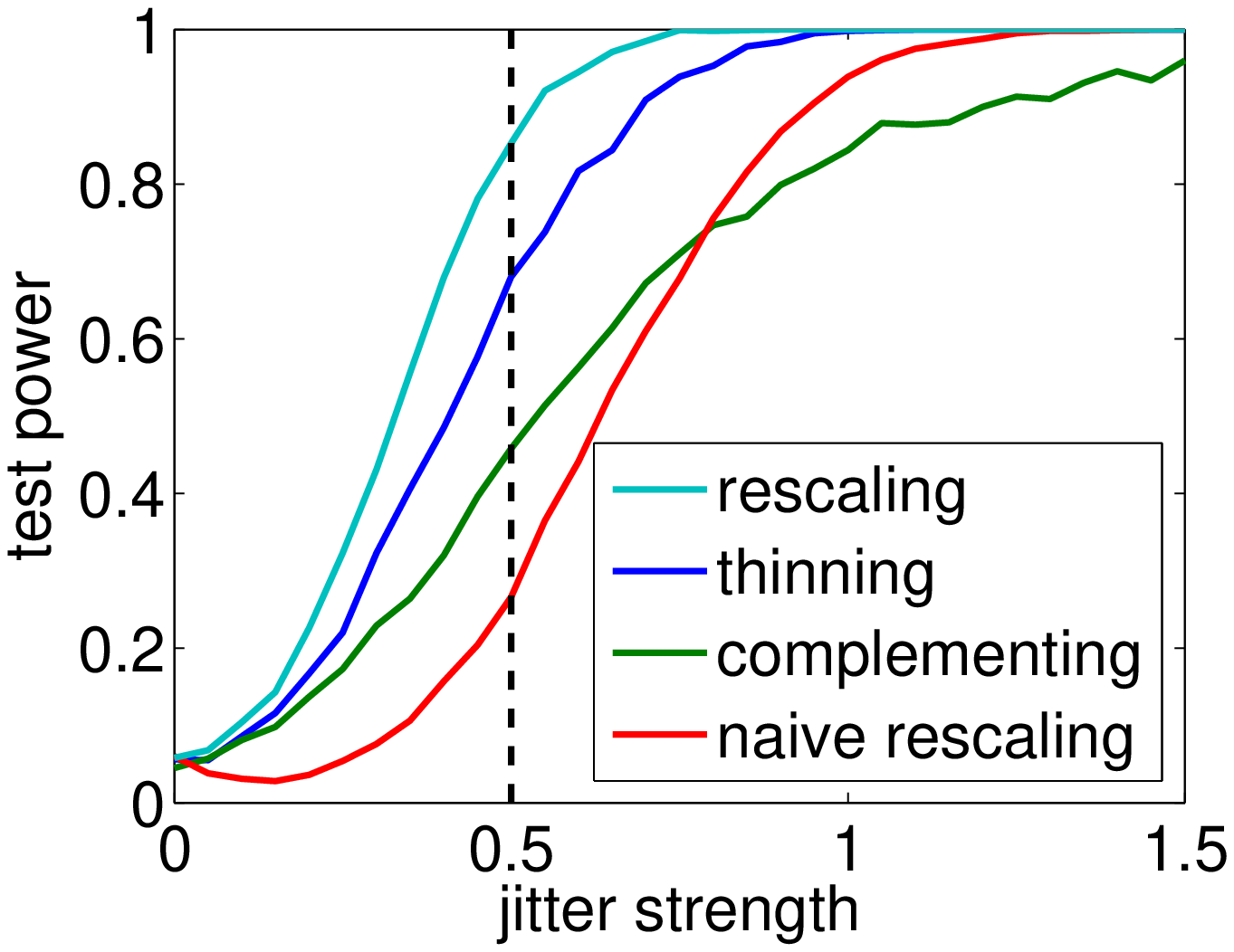}}
 \subfloat[ROC curve]{\includegraphics[width=0.33\linewidth]{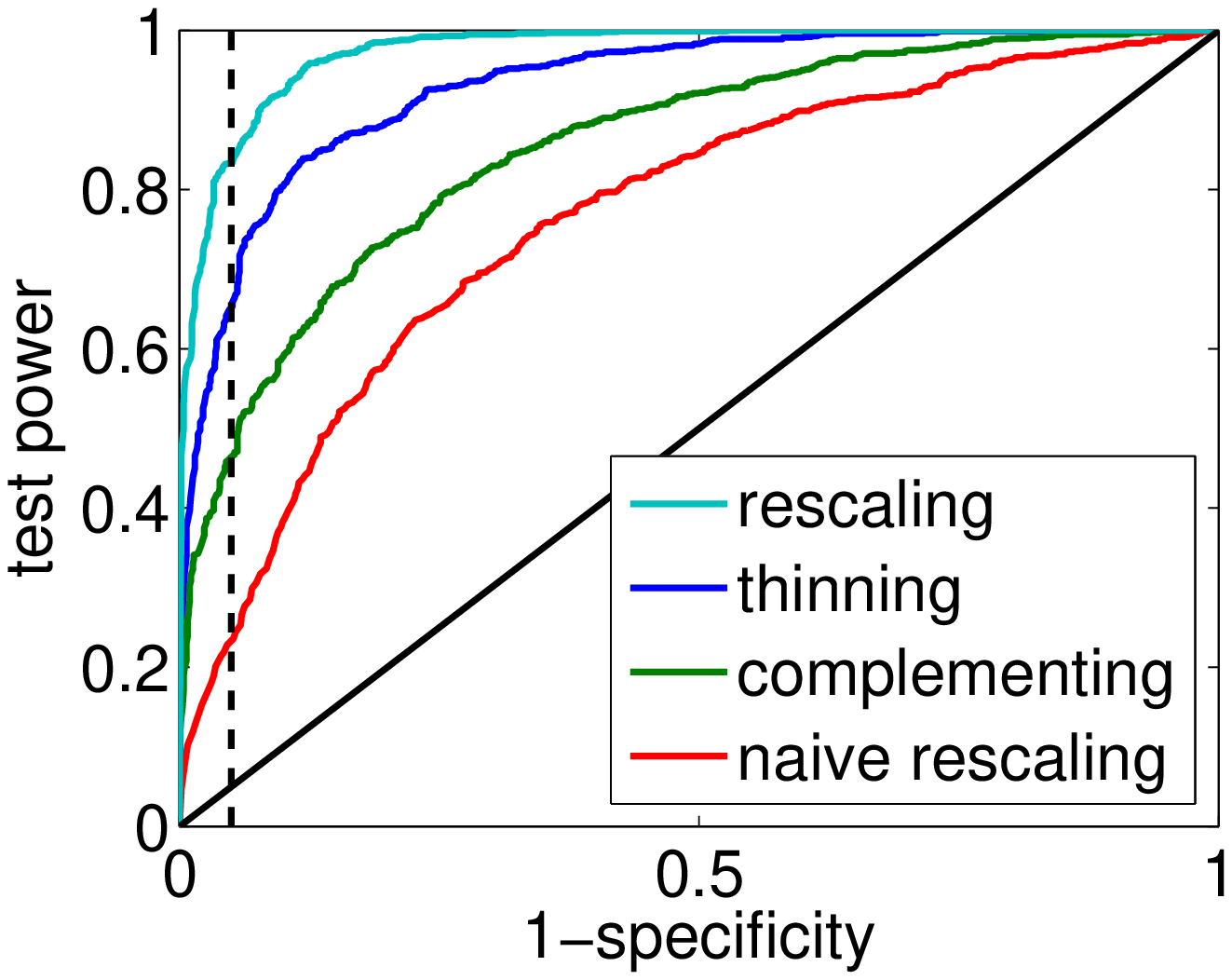}}
 \caption{Renewal process. (A) Inter-spike interval distributions for the undistorted (black line) and distorted models (medium jitter, $\beta = \exgammamedjitter$ and strong jitter, $\beta = \exgammahighjitter$). For comparison, a sample ISI histogram from one of the simulations is shown in gray. Note that the mean of the three distributions is matched to be the same (vertical dashed line). (B) The test power of each test as a function of the jitter strength. The dashed line indicates the level of the medium jitter strength (red line in figure A). (C) ROC curve analysis for an intermediate jitter strength of $\beta = \exgammamedjitter$. The intersection of the curves with the dashed line corresponds to the test power at a significance level of $\alpha = .05$.}
\label{fig:exgamma}
\end{figure}

The spike train was generated from a true model, following a Gamma distribution with scale parameter $A = \exgammatruescale$ and shape parameter $B = \exgammatrueshape$: $p(\Delta t) = (\Delta t)^{B-1} \frac{e^{-\frac{\Delta t}{A}}}{A^{B}\Gamma(B)}$. Wrong models were generated by scaling the shape and scale parameter by a factor of $1 + \beta$ ("jitter") while keeping the expected value of the distribution constant (i.~e.~$B' = (1+\beta) B$, $A' = (1+\beta)^{-1} A$) (Figure~\ref{fig:exgamma}A). For each jitter strength, $N = \exgammaNrep$ data sets of length $T = \exgammaT~s$ were generated from the true model and the wrong model and the tests were applied.

The analysis of test power for each test and the ROC curve analysis for an intermediate jitter strength reveal that time-rescaling is slightly superior to thinning and complementing (Figure~\ref{fig:exgamma}B and C). The \naive time-rescaling performs worst (adjusted significance level for the KS test, $\alpha = .017$).

\subsection{Example: Inhomogeneous Spike Response Model}

\newcommand{\exsrmT }{20 } 
\newcommand{\exsrmNrep }{1000 }
\newcommand{\exsrmmedjitter }{0.4 }
\newcommand{\exsrmhighjitter }{1.0 }
\newcommand{\exsrmncoeff }{40 }
\newcommand{\exsrmbandlimit }{1 }


We model an inhomogeneous spike response model with escape noise using a Bernoulli-GLM \cite{Gerstner2002Book}. The spiking probability is modulated by an inhomogeneous rate $r(t)$. Additionally, for each spike, a post-spike kernel is added to the process intensity. The rate function is modeled like in the first example as a band-limited function $r_{t_i} = r(t_i) = \sum_{j=1}^{J = \exsrmncoeff} u_j \frac{sin(2 \pi f (t_i-\frac{j}{J}T))}{\pi (t_i-\frac{j}{J}T)}$ with $f = \exsrmbandlimit$~Hz and $J = \exsrmncoeff$ coefficients that were randomly drawn from a uniform distribution on the interval $[-0.2,0.2]$. The post-spike kernel $\eta(\Delta t)$ is modeled as a sum of three exponential functions ($\tau = 5$~ms, $25$~ms and $1$~s) with appropriate amplitudes as to mimick a relative refractory period, a small rebound and a slow (inhibitory) adaptation. To construct the Bernoulli-GLM, the spiking probability $p_i$ per bin of length $\Delta = 1$~ms is $p_i = \frac{1}{1+exp(-s_i)}$ with $s_i = -3 + r_{t_i} + \sum_{\{u_j\}<t_i} \eta(u_j-t_i)$.

\begin{figure}
\centering
 \subfloat[intensity function]{\includegraphics[width=0.33\linewidth]{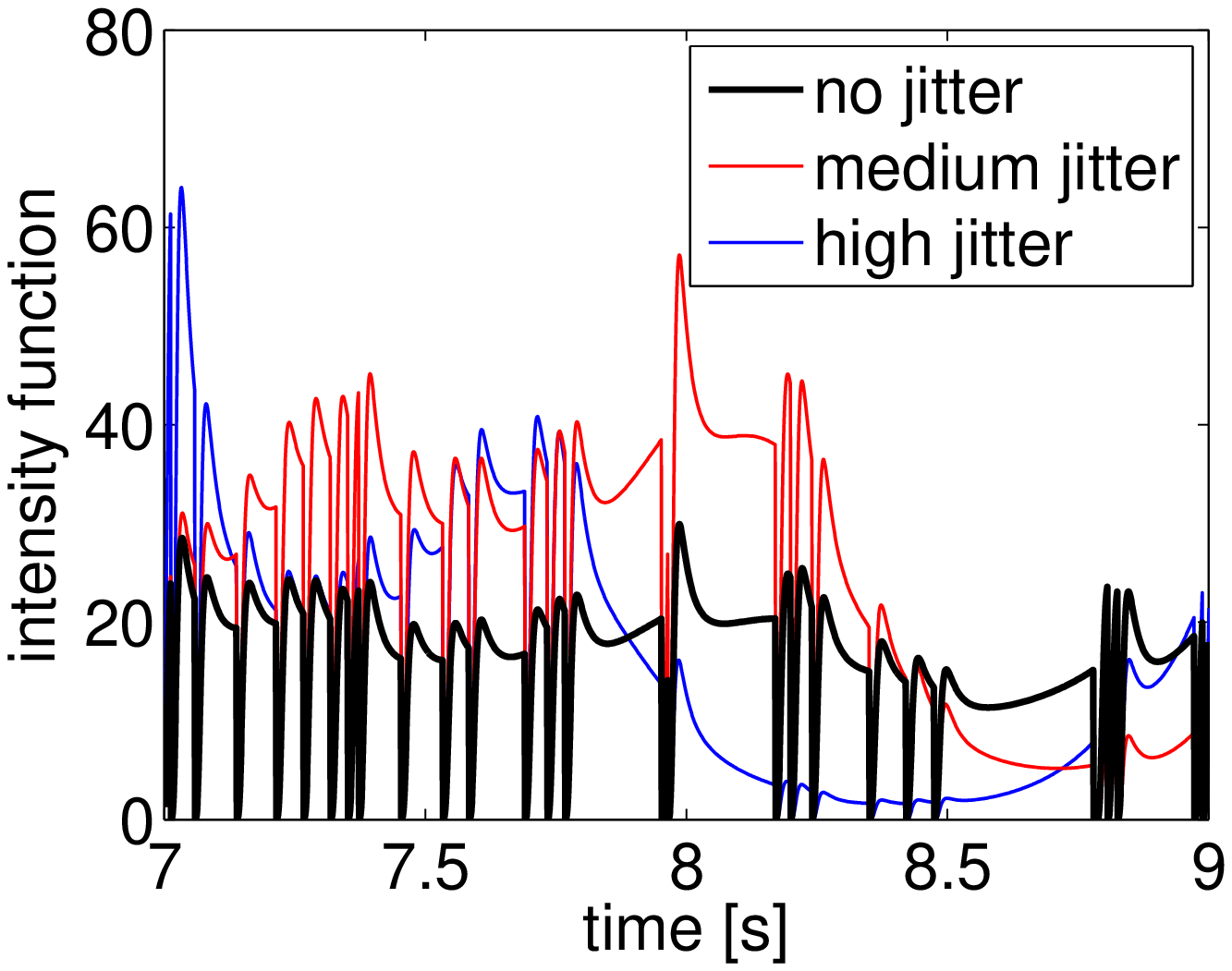}}
 \subfloat[test power]{\includegraphics[width=0.33\linewidth]{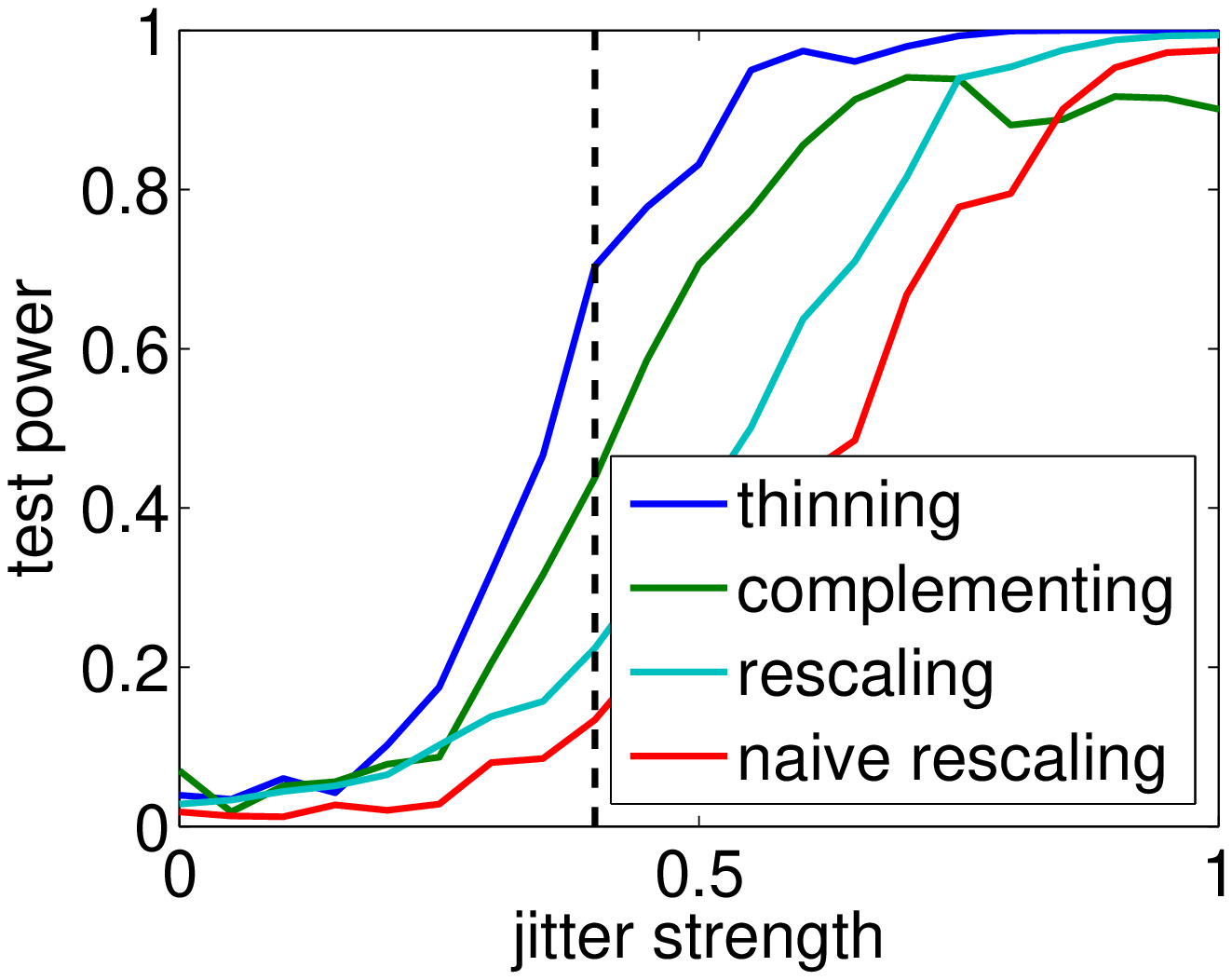}}
 \subfloat[ROC curve]{\includegraphics[width=0.33\linewidth]{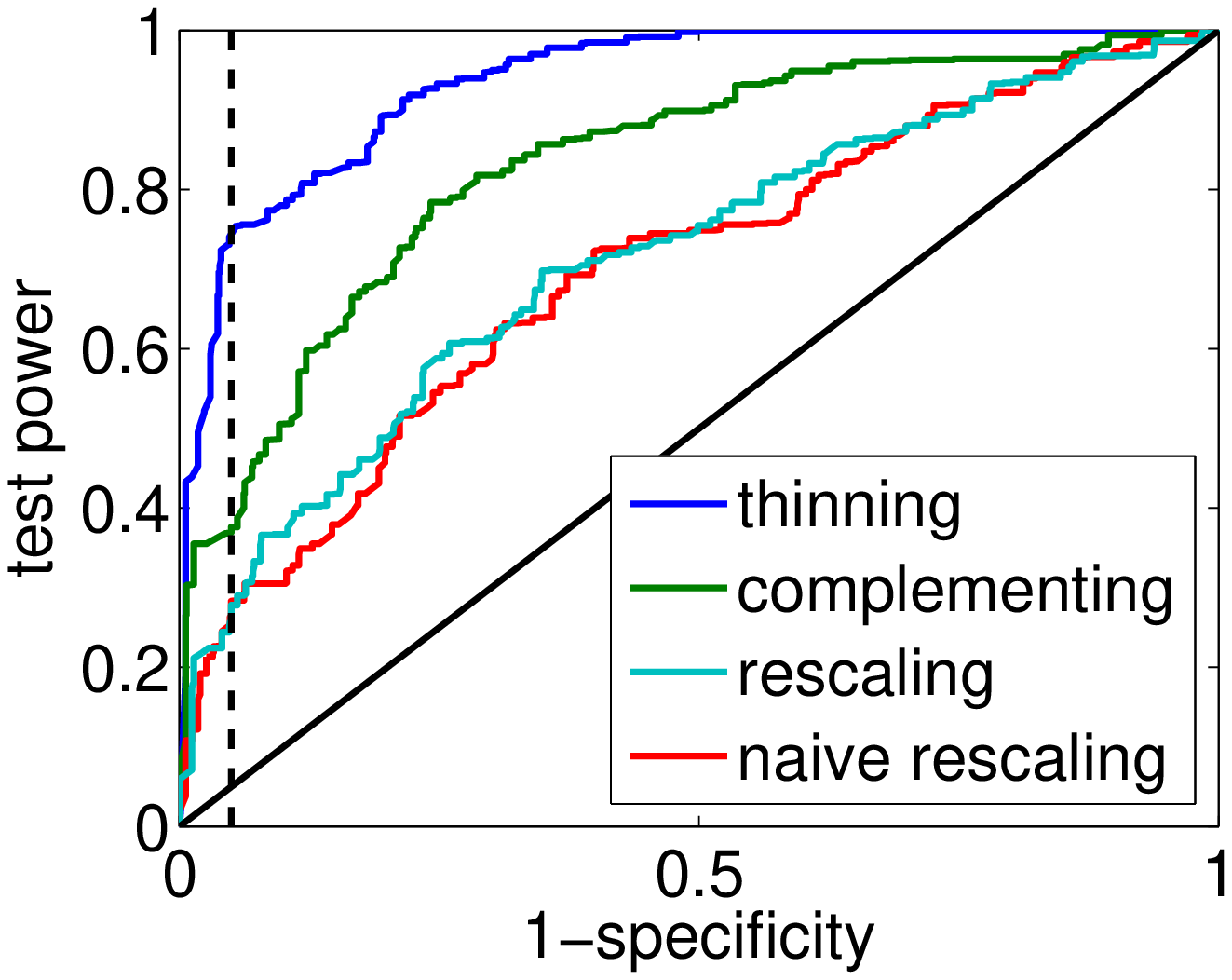}}
 \caption{Inhomogeneous Spike Response Model. (A) Sample intensity functions for an undistorted intensity (black line) and two misspecified models (medium jitter, $\beta = \exsrmmedjitter$ and strong jitter, $\beta = \exsrmhighjitter$). (B) The test power of each test as a function of the jitter strength. The dashed line indicates the level of the medium jitter strength (red line in figure A). (C) ROC curve analysis for an intermediate jitter strength of $\beta = \exsrmmedjitter$. The intersection of the curves with the dashed line corresponds to the test power at a significance level of $\alpha = .05$.}
\label{fig:exsrm}
\end{figure}

A binary time series (the spike train) was generated for a duration of $T = \exsrmT$~s. The jittered models were constructed by adding a jitter $\beta$ on the coefficients of the inhomogeneous rate modulation (Figure~\ref{fig:exsrm}A). For each jitter strength, $N = \exsrmNrep$ data sets were generated from the true model and the wrong model and the tests were applied.

Both thinning and complementing are able to detect smaller distortions than both the time-rescaling on the surrogate and discrete data (Figure~\ref{fig:exsrm}B, adjusted significance level for the \naive rescaling, $\alpha = .018$). A ROC curve analysis for an intermediate jitter strength ($\beta = \exsrmmedjitter$) supports this finding (Figure~\ref{fig:exsrm}C).

\section{Discussion}
\label{sec:discussion}

Assessing goodness-of-fit for Generalized Linear Models has mostly been done by applying the time-rescaling transformation that is defined for point processes, assuming a match between those approaches. When the per-bin probability of spiking cannot be regarded as low, this approximation breaks down and creates a bias when applying the time-rescaling transformation \cite{Haslinger2010DTR}. In a first step, we proposed a procedure to create surrogate point processes from discretized models, such as Bernoulli- and Poisson-GLMs, that do not exhibit this bias. Throughout all the examples, the time-rescaling theorem applied to the surrogate point process was systematically better than applying the \naive time-rescaling on the discrete data. Since only the adjusted time-rescaling procedure allows to reliably control the specificity of the test, it should be preferred over the classical time-rescaling in all cases where discretized models are used.

We have presented two alternatives to an application of the time-rescaling theorem: For the first procedure, the observed spike train is thinned according to the value of the conditional intensity at the time of spikes. The resulting process is then a homogeneous Poisson process with a rate that is equal to the lower bound on the conditional intensity. The second proposed method builds on the idea that an intensity function $\lambda(t)$ with an upper bound $C$ can be filled up to a homogeneous Poisson process of rate $C$ by adding spike samples from the complementary process $C-\lambda(t)$. The proposed tests work best if the lower and upper bounds are tight. However, in most practical cases, especially the lower bound will be prohibitively low to apply any statistical test on the thinned process. As a remedy, we proposed to consider only regions of $\lambda(t|H(t))$ for which the intensity exceeds a given threshold and repeat the thinning for different thresholds. This successfully overcomes the limitation that may have -- up to now -- prevented the use of the thinning algorithm as a goodness-of-fit measure for neural models.

The three tests are complementary in the sense that they are sensitive to different deviations of the modeled and true intensity function. Time-rescaling is only sensitive to the total integral of the intensity function between spikes, while thinning exclusively considers the intensity function at the time of spikes and is insensitive to its value at places where no spikes occurred. Complementing is sensitive to the exact shape of $\lambda(t)$ regardless of where the spikes from the original observations are.

For the examples of an inhomogeneous Poisson process and the Spike Response Model, thinning and complementing outperform the sensitivity of the simple time-rescaling procedure. They can detect deviations from the model that are only half as large as the ones necessary to alert the test based on time-rescaling. For modeling renewal processes, time-rescaling was slightly advantageous compared to the to other methods. This should not come as a surprise since the time-rescaling test is known to be sensitive to modeling the distribution of inter-spike intervals \cite{Brown2002}.

Beside from likelihood criteria \cite{Pillow2009,Wood2005NIPS,Berkes2008NIPS}, there exist few goodness-of-fit tools for neural models based on Generalized Linear Models \cite{Ogata1981,Brown2003}. With the proposed procedure for surrogate point processes, we bridge the gap between such discrete models and point processes. That enables to make use of additional tests from this domain, such as thinning and complementing procedures. We expect these to be valuable contributions to the general practice of statistical evaluation in modeling single neurons as well as neural populations.

\subsubsection*{Acknowledgments}
Felipe Gerhard thanks Gordon Pipa and Robert Haslinger for helpful discussions. Felipe Gerhard is supported by the Swiss National Science Foundation (SNSF) under the grant number 200020-117975.

\newpage

\small{

}

\end{document}